# Real Time Industrial Monitoring System

Rahul D. Chavhan[#1], Sachin U. Chavhan[#2], Ganesh B. Chavan[#3]

[#1]*Assistant Professor, Department of Electronics & Telecommunication, SGBA University, Amaravati
JCOET, Arni Road, Kinhi, Yavatmal, Maharashtra, INDIA*

[#2]*Assistant Professor, Department of Electronics & Telecommunication, University of Pune, Pune
SF's SIEM, Maihiravani, Nashik, Maharashtra, INDIA*

[#3]*Assistant Professor, Computer Engineering, North Maharashtra University, Jalgaon
GF's GCOE, Bhusaval Road, Jalgaon, Maharashtra, INDIA*

*Abstract*— **Industries are the biggest workplace all over the world, also there are large number of peoples involves as a worker and most of them are work as a machine operator. There are many systems developed for industrial work place, some of them, monitors machine processes and some do monitoring and control of machine parameters. Such as speed, temperature, production batch count etc. However there is no such system available that provides monitoring of operator during their work is in progress at workplace. This paper proposes the monitoring of the operators and the machines, by Real time Operator -Machine Allocation and monitoring system (Omams). Omams allocates a work machine to worker at entry point itself. It uses automation with RFID and one of the standards of wireless communication method. The system can be industry specific. Through this research paper our approach is to make fair allocation of machine to the operator in industry and reduce hassle for efficiency calculations.**

## I. INTRODUCTION

The focus of proposed system is real time monitoring of operator in workshop which is necessary for any industrial management. In recent years the focus was on machine parameter monitoring; now we are trying to design a system which is beneficial to the operators and management personnel by providing them information about machine utilization.

### A. Industrial Operation

In any industry, there are many departments. Each department may have many separate workshops. For each workshop, there may be maximum three shifts in a day. During the shift change, operators working in the workplace are relieved and entering the workplace is allocated the machines. In this case study we are trying to provide automated allocation and monitoring of operator at work place.

Normally every industry keeps two entries of the workers. First is at the entry gate of company (i.e. the time office or Personnel department) and second is at workshop supervisor register of individual worker. One is kept for salary and work hours count and later is kept for production floor monitoring or efficiency monitoring.

### B. Radio Frequency Identification

Radio-frequency identification (RFID) is the wireless non-contact use of radio-frequency electromagnetic fields to transfer data, for the purposes of automatically identifying and tracking tags attached to objects. The tags contain electronically stored information. Unlike a bar code, the tag does not necessarily need to be within line of sight of the reader, and may be embedded in the tracked object [1]. The *radio frequency identification (RFID)* is the technology similar in theory to bar code identification. With RFID, the electromagnetic or electrostatic coupling in the RF portion of the electromagnetic spectrum is used to transmit signals. An RFID system consists of an antenna and a transceiver, which read the radio frequency and transfer the information to a processing device, and a transponder, or tag, which is an integrated circuit containing the RF circuitry and information to be transmitted. The key differences between RFID and bar code technology are RFID eliminates the need for line-of-sight reading that bar coding depends on. Also, RFID scanning can be done at greater distances than bar code scanning. High frequency RFID systems (850 MHz to 950 MHz and 2.4 GHz to 2.5 GHz) offer transmission ranges of more than 90 feet, although wavelengths in the 2.4 GHz range are absorbed by water (the human body) and therefore has limitations. RFID is also called *dedicated short range communication (DSRC) [2]*. An RFID reader's function is to interrogate RFID tags. The means of interrogation is wireless and because the distance is relatively short; line of sight between the reader and tags is not necessary. A reader contains an RF module, which acts as both a transmitter and receiver of radio frequency signals [3].

## II. RELATED WORK

In the current scenario of Industry, the research work was carried on monitoring and control of parameters such as real time production monitoring; a production monitoring is a system that is used in real time to record production line problems. It is constructed using programmable logic controller and sensors to collect data from production lines [4], real time monitoring of an industrial batch process, the process shares many similarities with other batch processes in that cycle times can vary considerably, instrumentation is limited and inefficient laboratory assays are required to determine the end-point of each batch [5], real time monitoring of complex industrial process with particle filters, the application of particle filtering algorithms to fault diagnosis in complex industrial processes [6], wireless monitoring and control of networks [7], wireless sensors for monitoring workers health, a wireless/mobile sensor framework: to estimate production efficiency through remote physiological monitoring in all conditions; to suggest effective work schedules according to individual physiological conditions; to propose safety guidelines for tradesmen; and to validate the production efficiency in the construction industry [8], adaptive real time monitoring for large scale networked system monitoring, it is continuous real-time monitoring, which is essential for the realization of adaptive management systems in large-scale dynamic environments. Real time monitoring provides the necessary input to the decision making process of network management [9], monitoring industrial energy and carbon flow [10], etc. The research work has either focus of technical parameter of machine/processes or of data and security monitoring of health and environment.

In our system we are monitoring the operator throughout working hours by allotting the operator to the specific machine by using different modules.

### III. PROBLEMS IN CURRENT SYSTEM

There are many problems and issues remains unresolved in current model of work distribution and work monitoring. To enlist a few are as below:
1. There should be one supervisor to record or monitor entry and exit time and one at job allocation point to allocate a duty to the operator.
2. The production floor supervisor is assumed as 100% fact known person for the production floor.
3. Less exposure of technology, for example supervisor allots all operators to duty; workers don't have any information about it.
4. Manual monitoring is required; no one understands machine utilization except supervisor.
5. There may be fake allotment of machine to operator by supervisor.
6. Two point registrations consume valuable work time and hence decreased efficiency.
7. Manual recordkeeping cannot be accurate all the time.

These problems are observed from case study on industrial shift change process. In proposed systems the management issue and technical issues are resolved by combination of two technologies, RFID, wireless network and automation software. It can be a great help for monitoring of operators, machines and management.

### IV. PROPOSED MODEL

Our proposed model is an attempt to solve above problems with the help of Real time industrial operator monitoring system. It uses RFID to identify the worker and automation of the system.

The system will have three main components:
1. Scan-In unit
2. Central unit
3. Scan- Out unit

One of the wireless standards methods can be used for connectivity between Scan-In/ Scan-Out Unit and Central Unit.

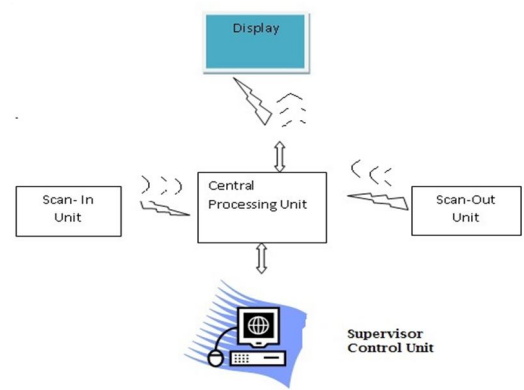

Fig. 1 Architecture of System

Through this research work we are attempting to propose model which minimizes the problem of allotment. Moreover transparency in the system comes with the use of technology. In this model we have enhanced the use of RFID with communication Network. The radio frequency identification (RFID) is unique identity for an employee, which can be easily detected with the help of RFID reader. Therefore we have suggested the use of embedded system for scanning and allotting the machine to the operator.

### V. METHODOLOGY

Our proposed model is divided in the following sub-module.
(A) In.
(B) Out.

(C) Allocation Procedure.

In an attempt with the Industry our proposed Model suggest the use of RFID in the employee card, which contains the code, for example A24564 is operator specific and identify the operator from whom the Code is extracted. A sample of RFID tag image is given in the following Figure 1.

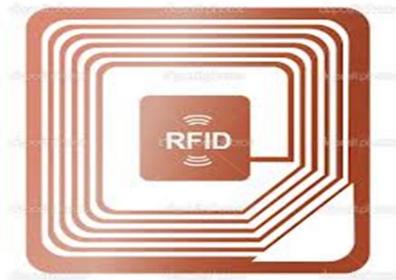

Fig. 2 Sample of a RFID on a Card.

So the steps in above proceedings in operator viewpoints are:-

A. *In*

The scan in unit serves two purposes in the system; first it will recognize the authenticated workmen for the industry. Second it will deliver a note or machine number for work along with job allocation.

Each employee has been given an identity card with a unique RFID.

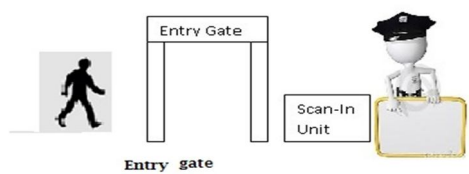

Fig. 3 Process of Check-In

1. Operator must produce his card at Scan-In unit to check-In. It facilitates faster scanning of employee code and avoids queuing at entry point.
2. Operator places his card in his hand near Scan-In unit to read RFID to get in.
3. A Scan-In unit reads the RFID over the card of operator and verifies it.
4. After In process the information is sent to the central unit for allotment.

B. *Out*

1. Operator goes to the Scan-Out terminal and scans its RFID to get out.

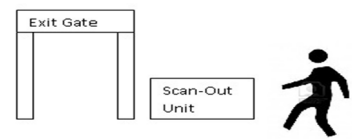

Fig. 4 Process of Check-Out

2. A Scan-Out unit reads the RFID over the card of operator and verifies it.
3. After Out process the information sent to the central unit for further process.

Once the Out procedure is done by the scan-out unit, the operator completes his work and leaves the workshop. But if the operator left the work in-between due to his personal reason then the scan-out unit will allow the operator to scan his RFID and it will send the information to central unit for allocation of machine to other operator.

C. *Allotment*

After getting information from scan-in unit and scan-out unit the central unit starts the procedure of allotment.
**1.** Get the list of operators who are checking-in from scan-in unit.
**2.** Get the list operators who are checking-out from scan-out unit.
**3.** Check the vacant machine list and allot the machine to the operators who are checking-in.
**4.** Check for operators who are not allotted and display a list of those with allotted operators list.
**5.** If in between any operator leaves the work, the machine he operating is allotted to the operator who is in waiting list.

For any monitoring application in industry, there are many parameters to be handled. In this system we are concentrating on machine operator. The operators are provided a unique code for this purpose. For monitoring operators and allotting them a specific machine, there are two registration units, one is at entry of industry i.e. Scan-In unit and other is in workshop i.e. Scan-Out unit. Both units are identical. These units may be located fixed at a specific location. Both the units scan the operator's code and allow them to enter and exit. Their unique numbers are then sent to the central unit continuously from both places i.e. Scan-In unit & Scan-Out Unit. The list from scan-in unit is of operators who are coming to work and from scan-out unit is of operators who are going from work. The central unit receives the data from both the units and allots a machine to each operator sequentially who check-in from scan-in unit and relieve the operator who check-out from scan-out unit, sends the corresponding information to display. There are two identical display units displaying the same information, one placed at workshop and other is at manager's office along with central unit. This cyclic process is continued till our requirement is fulfilled during shift change.

Whenever the number of operators exceeds the number of machines, central unit displays a waiting the list of remaining operators on the display. If numbers of operators are less than number of machine, then central unit displays the list of vacant machine. If in between any operator left the machine, due to some problem, he has to scan his card at Scan-Out unit for check-out at workshop. After check-out, Scan-Out unit sends this information to the central unit and then central unit assigns a next operator waiting for work by displaying this information at display unit automatically. The system provides a complete monitoring of operators and makes it more transparent.

### VI. BENEFITS

1. This technology will replace the complex shift changing procedure that is often carried by the industry.
2. Induction of this technology will facilitate supervisor to allot vacant machine operators.
3. It will enable operator to update the status of his turning up for the work.
4. After this updating, central unit will come to know about the machine of absent operator and will allot those machine waiting operator informing them by displaying or an SMS.
5. After allotting the machines to all the operators, if some machine still remains vacant then it will be reflected as available machine on the display and it could be allotted to operator which is willing to work for that machine from any of the workshop.
6. It attempts to reserve each and every machine even vacant for an hour.
7. It maintains the transparency in machine allocation and makes the shift changing process fast.

### VII. EXPERIMENTAL RESULTS

For experiment and trials, the scan in and scan out units are designed with LPC2148 ARM7TDMI processor, the central processing unit is designed in same manner.

In scan and central unit, we used interfacing of GSM SIM900 module for communication between scan unit and central unit. It uses GPRS for data sharing which provides secure data sharing between these units. The scan unit is also using the RFID reader for identifying the operator and scanning operator id.

SIM900 is a quad-band GSM/GPRS engine that works on frequencies GSM 850MHz, EGSM 900MHz, DCS 1800MHz and PCS 1900MHz. It gives the GPRS data downlink transfer: max. 85.6 kbps and GPRS data uplink transfer: max. 42.8 kbps speed. For communication between units, GSM wireless method had been implementing and gives security and required speed of data transfer with satisfying results. Fig. 2, Fig. 3 and Fig. 4 shows the experimental setup performed.

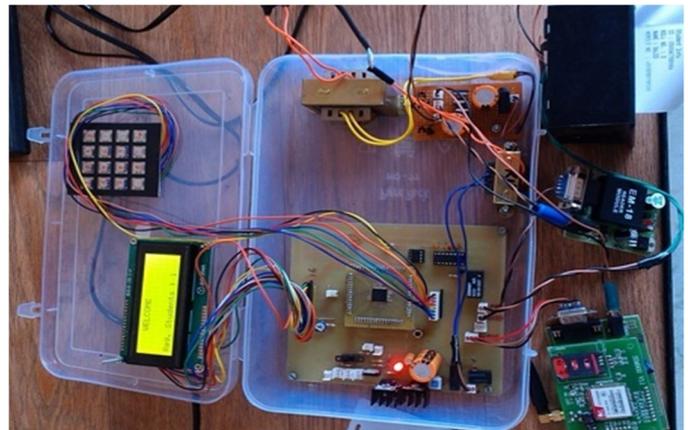

Fig. 5 Scan-In Unit

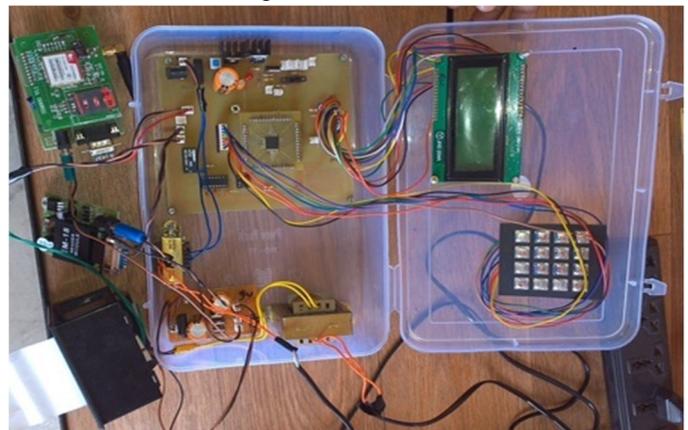

Fig. 6 Scan-Out Unit

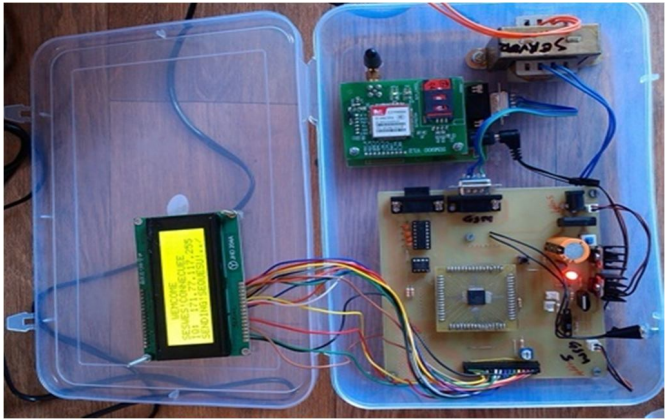

Fig.7 Central Unit

## VIII. CONCLUSION

This model proposes radical change in industrial operation and operator experience. Scan-in units are provided to workshop supervisor for smooth and faster scanning and verification of operators. RFID is embedded in the cards and this RFID is scanned by scan-in unit and scan-out unit. In RFID, an operator specific code is stored. When scanning unit encodes this code by In process or Out process, it redirects to central unit. In process updates the information of all operators available in the industry and let the central unit to make the machine reserve or vacant. Central unit allot the machines operators and if still some machines remain vacant then reflect them as available across industry from where any operator willing to work on it and can take that machine. Apart from this In, Out and allotment process is also provided to the operators. Out process provides the operators to break his work at any situation by Out procedure and at the same time, his vacant machine are provided to a waitlisted operator. These interfaces provide capability to work at any machine within the industry. These technology inclusions in the industry bring transparency and reduce the hassles of workers while shift change.

ACKNOWLEDGEMENT

The authors would like to thank Dr. K. P. Rane, H.O.D. E&TC Dept., Godavari Foundations Godavari COE, Jalgaon, for their kind support and Dr. B. K. Mukharjee, Principal, Godavari Foundation's Godavari College of Engineering, Jalgaon, for his encouragement and inspiration to work on this current research topic. They also would like to thank Prof. P. V. Phalak, Vice Principal, GF's GCOE, Jalgaon, for making software required and laboratory available.